\documentclass{ws-procs961x669}            

\begin{document}
\title{Transverse Force Imaging}

\author{Fatma P. Aslan}

\address{Department of Physics, University of Connecticut, Storrs, CT 06269, U.S.A.\\
$^*$fpaslan@jlab.org
}

\author{Matthias  Burkardt}

\address{Physics Department, New Mexico State University,\\
Las Cruces, NM 88003, U.S.A.\\
$^*$E-mail: burkardt@nmsu.edu}

\author{Marc Schlegel}

\address{Physics Department, New Mexico State University,\\
Las Cruces, NM 88003, U.S.A.\\
$^*$E-mail: schlegel@nmsu.edu}

\begin{abstract}
While twist-2 GPDs allow for a determination of the distribution of partons on the transverse plane, twist-3 GPDs contain quark-gluon correlations that provide information about the average transverse color Lorentz force acting on quarks. As an example, we use the nonforward generalization of $g_T(x)$, to illustrate how twist-3 GPDs can  provide transverse position information about that force.

\end{abstract}


\bodymatter

\section{Transverse Imaging}\label{aba:sec1}
For a transversely localized nucleon state, such as (${\cal N}$ is a normalization factor)
\begin{equation}
|{\bf R}_\perp =0,p^+,\Lambda\rangle \equiv
{\cal N} \int d^2{\bf p}_\perp |{\bf p}_\perp ,p^+,\Lambda\rangle,
\end{equation}
which has its transverse center of longitudinal momentum at the (transverse) origin, one can define transverse charge distributions as
\begin{eqnarray}\rho_{\Lambda^\prime\Lambda} ({\bf b}_\perp) 
&\equiv& \langle {\bf R}_\perp =0,p^+\!,\Lambda^\prime| \bar{q}({\bf b}_\perp)
\gamma^+ q({\bf b}_\perp)
|{\bf R}_\perp =0,p^+\!,\Lambda\rangle
\label{eq:F}\\
&=&\left|{\cal N}\right|^2 
\int\! \!\!d^2{\bf p}_\perp\!\!\!\int \!\!\!d^2{\bf p}_\perp^\prime
\langle {\bf p}_\perp^\prime,p^+\!,\Lambda^\prime| \bar{q}(0)
\gamma^+q(0)
|{\bf p}_\perp,p^+\!,\Lambda\rangle e^{i{\bf b}_\perp\cdot({\bf p}_\perp
-{\bf p}_\perp^\prime)}
\nonumber\\
&=& \int d^2{\bf \Delta_\perp} F_{\Lambda^\prime \Lambda}(-{\bf \Delta_\perp^2}) e^{-i{\bf b}_\perp \cdot {\bf \Delta_\perp}}.\nonumber
\end{eqnarray}
Here $\Lambda$, $\Lambda^\prime$ are the polarization of the target states, and $F_{\Lambda^\prime,\Lambda}$ is a superposition of the Dirac and Pauli form factors - details are depending on the polarizations. Note that in the $2^{nd}$ step in Eq. (\ref{eq:F}) it was crucial that the matrix element only depends in the $\Delta_\perp$, but not on the overall ${\bf P}_\perp=\frac{1}{2}({\bf p}_\perp+{\bf p}^\prime_\perp)$ - otherwise it would not be possible to factor out the ${\bf P}_\perp$ integration and cancel it against $|{\cal N}|^2$.

As a side remark, when one tries to localize a state in 3 dimensions, a factorization of the ${\vec P}$-integration is not possible due to various relativistic factors. As a result, a similar procedure in 3 dimension fails and the physical interpretation of 3-dimensional Fourier transforms of form factors as charge distribution in position space is flawed due to relativistic corrections when one looks at details smaller than the Compton wavelength of the target.

Very similar steps can be repeated for $x$-dependent distributions resulting in the position space interpretation for Generalized Parton Distributions (GPDs) \cite{mb:GPD}.

\section{Transverse Force}
The $x^2$ moments of the genuine twist-3 part of twist-3 PDFs are related to forward matrix elements of quark-gluon-quark correlations. 
For example, the polarized twist 3 PDF $g_2(x)$ can be cleanly separated from the leading twist contribution $g_1$ by measuring
the longitudinal (beam) - transverse (target) double-spin asymmetry in DIS. After subtracting the Wandzura-Wilczek contribution
$g_2^{WW}(x) \equiv -g_1(x) +\int_x^1 \frac{dy}{y}g_1(y)$, one is left with the genuine twist 3 part $\bar{g}_2(x)=g_2(x)-g_2^{WW}(x)$ (here we neglect quark mass contributions), whose $x^2$ moment reads \cite{qGq}
\begin{equation}
d_2\equiv 3\int dx\,x^2{ \bar{g}_2(x)} =
\frac{1}{2M{P^+}^2S^x} \left\langle P,S \left|
\bar{q}(0) \gamma^+gG^{+y}(0)q(0)\right|P,S\right\rangle
\label{eq:qFq}
\end{equation}
To understand the physical meaning of this correlator, we decompose the light-cone component of the gluon field strength tensor appearing in (\ref{eq:qFq}) in terms of color electric and magnetic fields 
\begin{equation}
\sqrt{2}G^{+y} = G^{0y}+G^{zy} = -{ E^y}+{ B^x}
=-\!\left({{\vec E}} + {\vec v}
\times {{\vec B}}\right)^y\end{equation}
for a quark that moves with the velocity of light in the $-\hat{z}$ direction - which is exactly what the struck quark does in a DIS experiment after having absorbed the virtual photon${\vec v}=(0,0,\!\!-1)$. Since the Gluon field is correlated with the quark density, this means that $d_2$ has the physical interpretation as the average color-Lorentz force acting on a quark in a DIS experiment right after (since the matrix element is local) having absorbed the virtual photon\cite{mb:force}. This is the same final state interaction (FSI) force that also
produces single-spin asymmetries.

\section{Transverse Force Imaging}
Since $d_2$ arises as the expectation value of a $\bar{q}Gq$ corellator in a plane wave state it can only provide volume-averaged information. In order to obtain position information, a momentum transfer must be involved. This is one of the motivations for
studying twist 3 GPDs. After subtracting Wandzura-Wilzek type terms and surface terms (terms that involve a twist 2 contribution multiplied by a momentum transfer),  $x^2$ moments of twist 3 GPDs allow determining non-forward matrix elements of $\bar{q}Fq$ correlators. This motivates parameterizing these matrix elements in terms of Lorentz invariant generalized form factors.

Lorents invariance implies that the matrix elements of $\bar{q}(0)\gamma^\rho G^{\mu \nu}(0)q(0)$ can be parameterized in terms of 8 generalized form factors \cite{abs2}

In the relevant case of $\bar{q}(0)\gamma^+ G^{+i}(0)q(0)$ this reduces to five form factors \cite{abs1}

\begin{eqnarray}\label{FormFactors}
\langle p',\lambda'|\bar{q}(0)\gamma^{+}igG^{+i}(0) q(0)|p,\lambda\rangle
&=&\overline{u}(p',\lambda')\Big\{\dfrac{1}{M^2}[P^{+}\Delta_{\perp}^i-P^{\perp}\Delta^{+})]\gamma^+\Phi_1(t)\\ \nonumber +\dfrac{P^+}{M}i\sigma^{+i}\Phi_2(t)
&+&\dfrac{1}{M^3} i \sigma ^{+\Delta}\big[P^+\Delta_{\perp}^i\Phi_3(t)-P^\perp \Delta^+\Phi_4(t)\big]\nonumber\\ 
&+&\dfrac{P^+\Delta^+}{M^3}i \sigma ^{i\Delta}\Phi_5(t) \Big\}u(p,\lambda)\nonumber
\end{eqnarray}
As was the case for twist 2 GPDs and charge form factors, a position space interpretation requires a vanishing longitudinal momentum transfer $\Delta^+=0$. 
The transverse Fourier transform of these generalized form factors have the following interpretation
\begin{itemize}
\item $\Phi_1$ describes an axially symmetric transverse force in an unpolarized target
\item $\Phi_2$ describes a force field perpendicular to the transverse polarization of the target, to a $\perp$ position resolved
Sivers force \cite{Sivers}
\item $\Phi_3$ describes a tensor type force similar to what one would expect from a color magnetic dipole field correlated with the target tranverse spin. 
\item $\Phi_4$ \& $\Phi_5$ involve a factor $\Delta^+=0$ and thus do not contribute to $\perp$ force tomography
\end{itemize}

\section{summary}

The Fourier transform of twist 2 GPDs w.r.t. the transverse momentum transfer provides transverse images of quark distributions. Taking $x^2$ moments of twist 3 PDFs allows determining the average transverse force that acts on a quark in a DIS experiments. Combining these two ideas, the transverse Fourier transform of the $x^2$ moment of twist 3 GPDs allows determining the spatially resolved transverse force that acts on a quark in a DIS experiment.

{\bf Acknowledgements:} 
M.B was supported by the DOE under grant number 
DE-FG03-95ER40965.  F.A. was partially supported by the DOE Contract No. DE- AC05-06OR23177, under which
Jefferson Science Associates, LLC operates Jefferson Lab.

\end{document}